\documentstyle[preprint,aps]{revtex}
\begin{document}
\draft
\title{Compressible $\nu=\frac{1}{2}$ state in a finite size study?}
\author{J.K. Jain}
\address{Department of Physics, State University of New York
at Stony Brook, Stony Brook, New York 11794-3800}
\date{February 24, 1994}
\maketitle

For a meaningful numerical
investigation of thermodynamic properties, systems bigger than the
correlation length need to be
studied. Fortunately, due to the existence of a gap,
the incompressible fractional quantum Hall
states have a finite correlation length, which, for some states
(e.g., 1/3, 2/5) happens to be smaller than the system sizes that
can be studied. On the other hand, since the Fermi liquid is
a critical state with power law correlations, one might expect that
finite system studies
are less likely to provide any conclusive results.  Rezayi
and Read (RR)\cite{RR} have recently claimed that their
numerical study provides ``convincing evidence for the correctness
of the Halperin-Lee-Read
\cite {HLR} theory of a compressible fermi-liquid-like state" at
$\nu=1/2$. We argue below that their study sheds little
light on the on the {\em thermodynamic} nature of the {\em
compressible} state at $\nu=1/2$.

RR have studied $N$-electron systems at flux $N_{\phi}=2(N-1)$
in the spherical geometry. Their numerical calculations test a
special case of the composite fermion theory \cite {Jain}, which provides
a general relationship between the low-energy spectra and eigenstates
of the systems at $N_{\phi}$ and  $N_{\phi}^*=N_{\phi}-2(N-1)$.
Extensive calculations of this type have
been performed in the past, {\em including for} $N_{\phi}=2(N-1)$
\cite{Dev,Wu,F1}. RR merely interpret the results differently.
For example, the system
with $(N,N_{\phi})=(9,16)$ was studied earlier \cite {Wu,Gros}  as an
incompressible state, but RR label it as the $\nu=1/2$ compressible
Fermi liquid state.
This raises the question of whether $N_{\phi}=2(N-1)$ is a credible finite
size representation of the compressible $\nu=1/2$ state.
To answer this, let us consider the incompressible states
at $\nu=n/(2n\pm 1)$. There are compelling reasons to believe that these
are described by finite systems with $N_{\phi}=2(N-1)\pm (N-n^2)/n$
\cite {Haldane,Gros,Dev}; {\em with no exception so far}, these systems
have uniform, non-degenerate and incompressible ground states;
moreover, for $n=1$ and 2 the gap has been found to vary smoothly
with $N$, which allows finite size scaling to be performed to
determine the gap in the thermodynamic limit \cite {Gros}.
The system $(N,N_{\phi}) =(9,16)$
is incompressible according to this equation \cite {F2}.
Indeed, the exact spectrum of this system (see Fig.3 of RR
\cite {F3}) looks like that of any other {\em incompressible} state
(for typical spectra, see \cite {Haldane}):
the non-degenerate ground state is well separated
from other states; there is an exciton branch;
and the separation between the ground state and the exciton branch
is $\approx$ $0.06 e^2/\epsilon l$, which is roughly of the same order
as the estimated energy gaps of the incompressible 3/7 and 3/5  states.
Similarly, $(N,N_{\phi})=(4,6)$ also represents an incompressible
state. These considerations  make  identification
of $N_{\phi}=2(N-1)$ with a {\em compressible} state implausible.
Indeed, as is clear from the RR work, the properties of the system at
$N_{\phi}=2(N-1)$ fluctuate wildly as a function of $N$ (at least
for small $N$), and no smooth extrapolation to
$N\rightarrow\infty$ is possible.

The difficulty in approaching 1/2 is also clear from the well
known (and frustrating) fact that the minimum number of
electrons required for the study
of the $n/(2n\pm 1)$ state increases with $n$. (This state
is related to the $n$-filled-Landau-level state, and
$N\geq n^2$ electrons are required to fill $n$ Landau levels in the
spherical geometry.) With
$N<15$, which is the present limit of exact diagonalization studies,
even 4/9 ($n=4$) is out of reach, to say nothing of 1/2, which is obtained
in the limit $n \rightarrow \infty$!

This work was supported in part by
the NSF under Grant No. DMR93-18739. Useful conversations with S.A.
Kivelson, X.G. Wu and F.C. Zhang are gratefully acknowledged.


\begin{thebibliography}{99}
\bibitem{F2} It could be identified as either 3/7 or 3/5.
\bibitem{F3} Note that the $L=0$ ground state (not shown in the
figure) is at zero energy.
\end{thebibliography}
\end{document}